\documentclass[english,aps,eqsecnum,superscriptaddress,nobibnotes,nofootinbib,preprint]{revtex4}
\usepackage[T1]{fontenc}
\usepackage[latin9]{inputenc}
\usepackage{xcolor}
\usepackage{pdfcolmk}
\definecolor{note_fontcolor}{rgb}{0.800781, 0.800781, 0.800781}
\usepackage{amsmath}
\usepackage{amssymb}
\usepackage{graphicx}
\PassOptionsToPackage{normalem}{ulem}
\usepackage{ulem}

\makeatletter

\newenvironment{lyxgreyedout}
  {\textcolor{note_fontcolor}\bgroup\ignorespaces}
  {\ignorespacesafterend\egroup}
\providecolor{lyxadded}{rgb}{0,0,1}
\providecolor{lyxdeleted}{rgb}{1,0,0}

\@ifundefined{textcolor}{}
{%
 \definecolor{BLACK}{gray}{0}
 \definecolor{WHITE}{gray}{1}
 \definecolor{RED}{rgb}{1,0,0}
 \definecolor{GREEN}{rgb}{0,1,0}
 \definecolor{BLUE}{rgb}{0,0,1}
 \definecolor{CYAN}{cmyk}{1,0,0,0}
 \definecolor{MAGENTA}{cmyk}{0,1,0,0}
 \definecolor{YELLOW}{cmyk}{0,0,1,0}
}

\usepackage{lscape}
\graphicspath{{Fig3/}}

\usepackage{babel}

\makeatother

\usepackage{babel}
\begin{document}

\title{Tracking the photodissociation probability of D$_{2}^{+}$ induced
by linearly chirped laser pulses}

\author{András Csehi }

\affiliation{Department of Theoretical Physics, University of Debrecen, H-4010
Debrecen, PO Box 5, Hungary}

\author{Gábor J. Halász }

\affiliation{Department of Information Technology, University of Debrecen, H-4010
Debrecen, PO Box 12, Hungary}

\author{Lorenz S. Cederbaum}

\affiliation{Theoretische Chemie, Physikalish-Chemisches Institut, Universität
Heidelberg, H-69120 Heidelberg, Germany}

\author{Ágnes Vibók }

\affiliation{Department of Theoretical Physics, University of Debrecen, H-4010
Debrecen, PO Box 5, Hungary}

\affiliation{ELI-ALPS, ELI-HU Non-Profit Ltd, H-6720 Szeged, Dugonics tér 13,
Hungary}
\begin{abstract}
In the presence of linearly varying frequency chirped laser pulses
the photodissociation dynamics of D$_{2}^{+}$ is studied theoretically
after ionization of D$_{2}$. As a completion of our recent work (J.
Chem. Phys. \textbf{143}, 014305 (2015)) a comprehensive dependence
on the pulse duration and delay time is presented in terms of total
dissociation probabilities. Our numerical analysis carried out in
the recently introduced light-induced conical intersection (LICI)
framework clearly shows the effects of the changing position of the
LICI which is induced by the frequency modulation of the chirped laser
pulses. This impact is presented for positively, negatively and zero
chirped short pulses.
\end{abstract}
\maketitle
\global\long\def\ket#1{\mbox{\ensuremath{|#1\rangle}}}
 \global\long\def\dket#1{\mbox{\ensuremath{|#1\rangle\rangle}}}
 \global\long\def\bra#1{\mbox{\ensuremath{\langle#1|}}}
 \global\long\def\dbra#1{\mbox{\ensuremath{\langle\langle#1|}}}
 \global\long\def\braket#1#2{\mbox{\ensuremath{\langle#1|#2\rangle}}}
 \global\long\def\expec#1{\mbox{\ensuremath{\langle#1\rangle}}}
 \global\long\def\R{{\sf R\hspace{1em}*{-0.9ex}%
\rule{0.15ex}{1.5ex}\hspace{1em}*{0.9ex}}}
 \global\long\def\N{{\sf N\hspace{1em}*{-0.9ex}%
\rule{0.15ex}{1.5ex}\hspace{1em}*{0.9ex}}}

\newpage{}

\pagenumbering{arabic} 

\section{Introduction \vspace{0.5cm}
}

The dynamics initiated in a molecule by photon impact is usually described
using the Born-Oppenheimer (BO) approximation ~\cite{born1}, where
the fast electrons are treated separately from the slow nuclei. In
this picture, electrons and nuclei do not easily exchange energy.
However, at some nuclear configurations, in particular, in the vicinity
of degeneracy points or conical intersections (CIs) this energy exchange
may become important ~\cite{horst,baer,graham,domcke,baer2,matsika,althorpe,bouakline}.
It is widely accepted today that conical intersections are fundamental
in the nonadiabatic processes which are ubiquitous in photophysics
and photochemistry ~\cite{domcke2,olivucci,robb}. At these CIs,
a molecule can switch efficiently between two different energy surfaces
on a time scale faster than a single molecular vibration. CIs are
associated with phenomena like dissociation, radiationless relaxation
of excited states, proton transfer, etc., and are also important in
biological molecules ~\cite{olivucci2,martinez2,sobolewski}.

For diatomic molecules that have only one degree of freedom, it is
not possible for two electronic states of the same symmetry to become
degenerate and as a consequence of the well-known noncrossing rule
an avoided crossing results. However, it is the situation only in
free space. It was presented in previous papers that conical intersections
can be created in a molecular system both by running and standing
laser waves even in diatomics ~\cite{nimrod,milan}. In this case
the laser light couples either the center of the mass motion with
the internal rovibrational degrees of freedom (in case of standing
laser field) or the vibrational motion with the emerged rotational
degree of freedom (in the case of running laser field) and the so-called
light-induced conical intersection (LICI) arises. By varying the laser
parameters (intensity and frequency), it is possible to affect the
position and structure of the LICIs, and thus control the dynamics
in order to study how molecules behave near a light-induced CI. An
interesting framework to understand this dynamics draws parallel with
conical intersections (CIs) between electronic surfaces in polyatomic
molecules.

A few years ago, we started a systematic study of the nonadiabatic
effects induced by laser waves in molecular systems. It has been demonstrated
along with other groups that LICIs have a significant impact on several
different dynamical properties (like molecular spectra, molecular
alignment, fragment angular distribution, branching ratio etc.) of
both diatomic ~\cite{gabor_1,gabor_2,gabor_3,gabor_4,chiang,pawlak}
and polyatomic molecules ~\cite{kim,philip,ware,corrales,ware2,sola,martinez}.
Recently we have made particular efforts for understanding the photodissociation
process of such a simple system as the D$_{2}^{+}$ in the LICI framework
~\cite{gabor_5,gabor_6,csehi,gabor_7,csehi_2,csehi_3}. The dissociation
of D$_{2}^{+}$ and its isotopologue, H$_{2}^{+}$ have been extensively
studied in the last couple of decades ~\cite{bandrauk_1,bandrauk_2,zavriyev,bandrauk_3,charron_1,atabek_1,charron_2,atabek_2,sanchez,sandig,atabek_3,posthumus,atabek_4,uhlmann,wang,anis_1,anis_2,hua,satrajit_1,satrajit_2,calvert,thumm,fischer_1,fischer_2,mckenna,he,furukawa,fischer_3,igarashi,haxton2,haxton}.
We note here that the effect of frequency chirped laser pulses on
the dissociation dynamics have been thoroughly investigated in the
last fifteen years ~\cite{krause,datta,kosloff,prabhudesai,natan,prabhudesai2,chang,forre,zhang,Natan,Novitsky}.
Several different dynamical properties like kinetic energy release
spectra, total dissociation probabilities as well as angular distributions
have been studied.

In our latest work ~\cite{csehi_2} we used linearly chirped laser
pulses for initiating the dissociation dynamics of D$_{2}^{+}$. In
contrast to the constant frequency (transform limited, TL) laser fields,
chirped pulses give rise to LICIs with a varying position according
to the temporal frequency change. In that work an interesting phenomenon
was revealed but the explanation remained unclear. Namely, it was
found that the amplitude of the periodic change in the dissociation
probability as a function of delay time is significantly compressed
by the chirped pulse compared to the transform limited situation.

The present investigation is a natural extension of our previous work
and now we provide a clear explanation for the origin of the physical
phenomena beyond the enhanced and suppressed dissociation processes
induced by chirped laser pulses. For that purpose, numerical simulations
are performed using different laser pulses. A full dimensional (2D)
scheme was applied to the problem in which the rotational angle is
assumed as a dynamic variable and, therefore, the LICI is explicitly
taken into account.

The article is arranged as follows. In the next two sections, we present
the methods and algorithms required for the theoretical study. The
calculated dynamical quantities and the numerical details are also
briefly summarized there. In the fourth section, we present and discuss
the numerical results for the D$_{2}^{+}$ system. In the last section
conclusions and future plans will be given.\newpage{}

\section{The D$_{2}^{+}$ molecule \vspace{0.5cm}
}

This section is devoted to the description of the most relevant features
of the system under investigation. The reason for choosing D$_{2}^{+}$
for the numerical analysis lies in its simplicity: (i) by having only
one electron in the system, electron correlation effects do not play
a role and (ii) by being a diatomic molecule there is only one internal
nuclear degree of freedom. All these properties make possible the
accurate treatment of D$_{2}^{+}$ from the theoretical point of view.
Furthermore the above simple properties allow us to purely focus on
physical aspects by excluding possible disturbing factors.

\vspace{0.5cm}
 
\begin{figure}[p]
\includegraphics[clip,width=8cm]{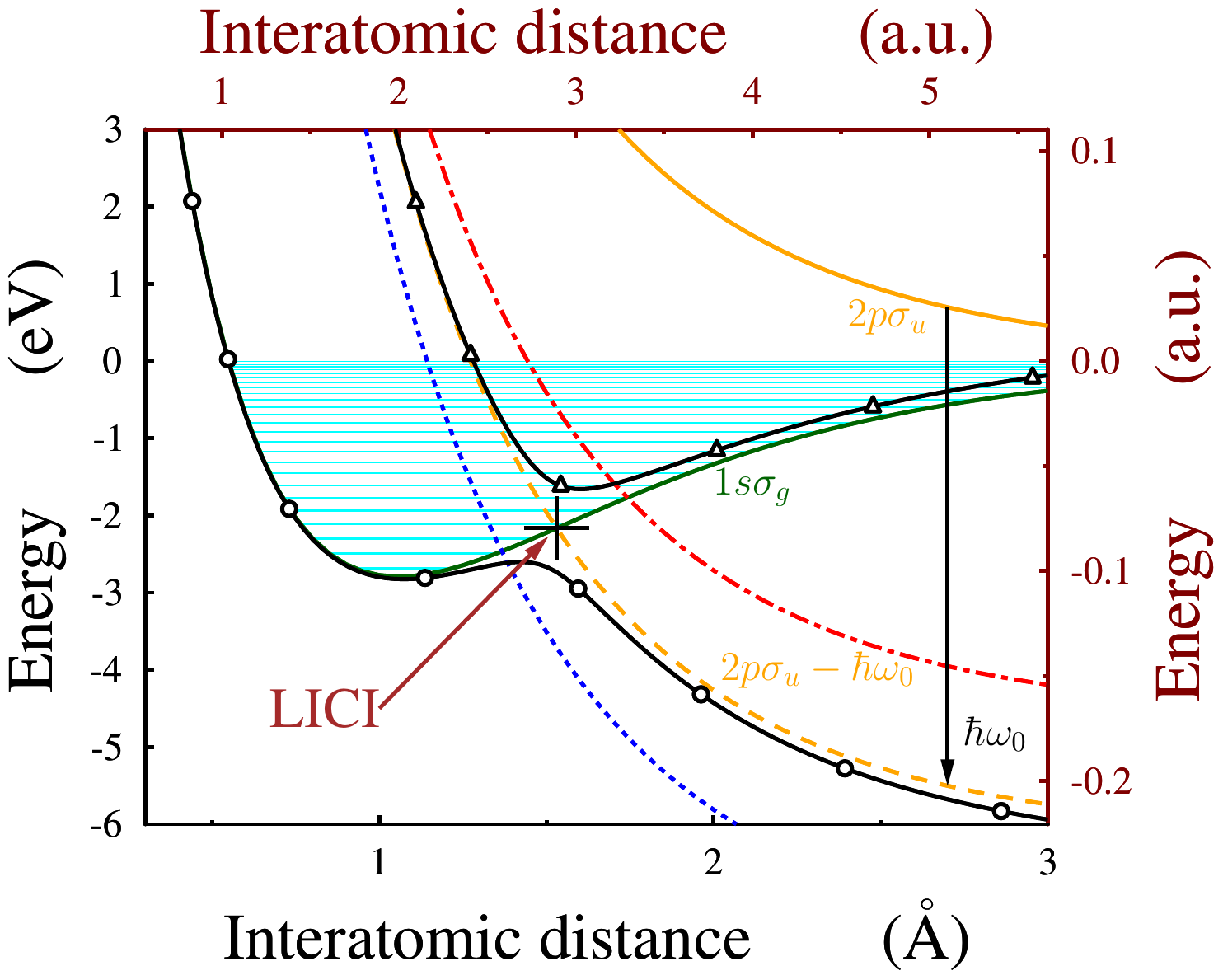} \caption{Potential energy curves of the D$_{2}^{+}$ molecular ion. Diabatic
energies of the ground $1s\sigma_{g}$ and the first excited $2p\sigma_{u}$
states are displayed with solid green and orange lines, respectively.
The field dressed excited states ($2p\sigma_{u}$ - $\omega_{0}$
) are presented for three different chirp situations (transform limited,
dashed orange line; positively chirped, dotted blue line; negatively
chirped, dashed-dotted red line). These field dressed excited states
($2p\sigma_{u}$ - $\omega_{0}$ ) form light-induced conical intersections
(LICIs) with the ground state. Only for the case of a transform limited
laser frequency$\omega_{0}=6.199\,eV$ and field intensity of $3\times10^{13}\frac{W}{cm^{2}}$,
a cut through the adiabatic surfaces at $\theta=0$ (parallel to the
field) is also shown by solid black lines marked with circles ($V_{l}$)
and triangles ($V{}_{u}$). For the transform limited case, we denote
with a cross the position of the LICI ($R_{LICI}=1.53\textrm{\AA}=2.891a.u.$
and $E_{LICI}=-2.16611eV$). }

\label{fig1} 
\end{figure}

In our treatment the ground and first excited electronic states of
D$_{2}^{+}$ molecular ion ($V_{1}=1s\sigma_{g}$ and $V_{2}=2p\sigma_{u}$)
are considered (see in Fig. 1). After a sudden ionization of the neutral
D$_{2}$ at $t=0$, its vibrational ground state is vertically transferred
to $V_{1}=1s\sigma_{g}$ and is considered as the initial wave packet
in our simulations. The vibrational wave packet created in this manner
is not in an energetic eigenstate of the $V{}_{1}$ diabatic potential,
it can be regarded as the superposition or Franck Condon distribution
of the vibrational eigenstates of D$_{2}^{+}$ .

After ionization, the $V{}_{1}(R)$ and the repulsive $V{}_{2}(R)$
electronic states are radiatively coupled by the laser field and light-induced
electronic transition occurs due to the non-zero transition dipole
moment $d(R)$ related to the electronic states:

\[
d(R)=-\bra{\xi_{1}^{el}}\sum_{k}r_{k}\ket{\xi_{2}^{el}}.\tag{2.1}
\]

In the space of the $V{}_{1}(R)$ and $V{}_{2}(R)$ electronic states
the time-dependent nuclear Hamiltonian including the ro-vibronic motion
and the light-matter coupling reads as follows (using atomic units
i.e. e=m$_{e}$=$\hbar$=1):
\[
H_{nuc}=(-\frac{1}{2\mu}\frac{\partial^{2}}{\partial R^{2}}+\frac{L_{\theta\phi}^{2}}{2\mu R^{2}}){\bf 1}+\begin{pmatrix}V_{1}(R) & -\epsilon(t)d(R)cos\theta\\
-\epsilon(t)d(R)cos\theta & V_{2}(R)
\end{pmatrix}.\tag{2.2}
\]
Here $R$ and ($\theta$ , $\phi$) are the vibrational and rotational
coordinates, respectively. $\theta$ is associated with the angle
between the molecular axis and the laser polarization direction and
the $\phi$-dependence is not present in our simulations as the initial
wave packet is in its rotational ground state ($J=0$). $L{}_{\theta\phi}$
denotes the angular momentum operator of the nuclei, $\mu$ is the
reduced mass and $\epsilon(t)$ is the electric field. 

The potential energy and the dipole moment functions were taken from
Refs. ~\cite{v1v2} and ~\cite{tdm}. The actual form of the applied
$\epsilon(t)$ electric field will be detailed in the next section.

After having all the necessary parameters of the system we are able
to carry out the nuclear dynamical simulations using the full time-dependent
Hamiltonian in (2.2). A very useful pictorial tool for understanding
the laser-matter interaction processes, especially in the strong field
physics is the so called Floquet representation of the nuclear Hamiltonian
\cite{gabor_1,gabor_4,Chu}. In our presentation we use this time-independent
scheme only for the illustration of the LICI phenomenon. In this picture
the $V_{2}(R)$ potential energy curve is shifted downwards by $\hbar\omega_{0}$
upon interaction with a laser field of $\omega_{0}$ frequency. As
a results of it a crossing is formed between the diabatic ground and
the dressed excited potential energy curves (Fig.1). After diagonalizing
the diabatic potential matrix in (2.2), the resulting adiabatic or
light-induced surfaces ($V{}_{l}$ and $V{}_{u}$ in Fig. 1) form
a conical intersection for geometry parameters determined by the following
conditions~\cite{nimrod,milan}:
\[
cos\theta=0\quad(\theta=\frac{\pi}{2})\qquad\mathrm{and}\qquad V_{1}(R)=V_{2}(R)-\hbar\omega_{0}.\tag{ 2.3}
\]

An important feature of the created light-induced conical intersections
as compared to the natural CIs is that their fundamental characteristics
can be modified by the external field. It has already been shown that
the intensity of the field determines the strength of the nonadiabatic
coupling, namely the steepness of the cone, while the energy of the
field specifies the position of the LICI. Therefore by applying chirped
or frequency modulated laser pulses the position of the LICI is continuously
changing during the dynamical process making possible in principle,
monitoring the time evolving nuclear wave packet, by properly chosen
chirp and laser parameters.


\section{Computational details \vspace{0.5cm}
}

In our numerical analysis carried out for D$_{2}^{+}$ the focus was
on the calculation of dissociation probabilities resulting from the
application of different external laser fields. This quantity was
derived according to the following formula ~\cite{mctdh_3}:

\[
P_{diss}=\int\displaylimits_{0}^{\infty}dt\bra{\psi(t)}W\ket{\psi(t)}\tag{3.1}
\]

where $-iW$ is the complex absorbing potential (CAP) applied at the
last $10\,a.u$. of the vibrational ($R$) grid.

The vibrational ground state of D$_{2}$ was taken as the initial
wave packet ($t<0\,fs$) for D$_{2}^{+}$ furthermore the molecule
was initially in its rotational ground state ($J=0$) in our calculations.
Linearly polarized Gaussian laser pulses were applied for initiating
the dissociation process. The dissociation probabilities were calculated
as a function of the delay time of the pulse with respect to the sudden
ionization ($t=0\,fs$). This delay time was described by the $t{}_{delay}$
parameter (the temporal maximum position of the field) which was ranging
from $0$ to $80\,fs$.

The central wavelength of the field was $\lambda_{0}=200\,nm$ ($\hbar\omega_{0}=6.199\,eV$)
and due to the chirping effect the frequency was changing linearly
around $t$$_{delay}$. The frequency changing rate depends on the
chirp parameter which is closely related to the pulse duration (as
will be detailed in the next section). In order to be able to follow
the nuclear motion we always tried to achieve proper chirp values.
Accordingly, maximum possible values had to be considered, namely
for $1$ $fs$ and $2$ $fs$ short pulses the chirp parameters were
$a$=$\pm0.3$ and $a$=$\pm0.15$, respectively, while in \cite{csehi_2}
for longer ($T$$_{p}$= $10\,fs$) pulses $a$=$\pm0.03$ were adopted.

Relatively large intensity ($10^{13}$$W/cm$$^{2}$) had to be utilized
so as to maximize the effect of the created LICI. Numerical simulations
will be presented with positively and negatively chirped as well as
transform limited (TL) laser pulses.

\subsection{The electric field \vspace{0.5cm}
}

Linearly polarized Gaussian laser pulses with linearly varying frequencies
have been applied during our time propagations. The electric field
is defined as the time derivative of the Gaussian shaped vector potential:
\[
\epsilon(t)=-\frac{\partial}{\partial t}A(t)\tag{3.1.1}
\]
\vspace{0.5cm}
 where the $A(t)$ vector potential has the following form: 
\[
A(t)=\frac{-\epsilon_{0}}{\sqrt[4]{1+\beta^{2}}}\cdot sin\left\{ \omega_{0}(t-t_{delay})-\frac{\alpha}{2}(t-t_{delay})^{2}+\varphi_{0}\right\} \cdot e^{-\frac{2log2}{1+\beta^{2}}\left(\frac{t-t_{delay}}{T_{p}}\right)^{2}}.\tag{3.1.2}
\]

In the above expression $\epsilon_{0}$ is the maximum field amplitude,
$\omega_{0}$ is the carrier frequency, while $T$$_{p}$ and $t$$_{delay}$
are the TL pulse duration (at FWHM) and the temporal maximum of the
pulse, respectively. In (3.1.2) $\varphi_{0}$ is the carrier envelope
phase (CEP, zero in the present study), $\alpha$ and $\beta$ are
chirp parameters closely related to each other: 
\[
\alpha=a\cdot\frac{\omega_{0}}{T_{p}}\frac{1}{(1+\beta^{2})}\tag{3.1.3}
\]
\[
\beta=\frac{a\cdot\omega_{0}\cdot T_{p}}{4\cdot log2}\tag{3.1.4}
\]
where the chirp parameter $a$ represents the relative change in $\omega_{0}$
within the pulse. As reflected by the expression of $A(t)$ (eq. (3.1.2))
the time dependence of the frequency is indeed linear: $\omega$$(t)$
= $\omega_{0}$ - $\alpha$ $(t-t_{delay})$. Here we note that other
forms can be considered as well, but the present study is concerned
with the above linear behavior. Our parametrization has the following
properties. $(i)$ The time integral of the field is zero (intrinsically
fulfilled by real pulses ~\cite{pulse}). $(ii)$ The spectrum of
the TL and the chirped pulses are the same. $(iii)$ The frequency
changing rate ($|\alpha|$) is maximal for the value of $|\beta|=1$.
This maximum is related to the TL pulse duration: $|\alpha|_{max}$
= $2log2/$$T$$_{p}^{2}$ . $(iv)$ Finally for small $\beta$ values
($\beta\ll1$) the frequency change is proportional to $a$ and $\omega_{0}$:
$\Delta\omega$ = $\alpha\cdot T_{p}\approx a\cdot\omega_{0}$ .

The first three of the above features are of great importance in practice,
as realistic chirped pulses exhibit the same kind of behavior.

\vspace{0.5cm}

\subsection{The propagation method \vspace{0.5cm}
}

For solving the time-dependent nuclear Schrödinger equation (TDSE)
of D$_{2}^{+}$ characterized by H$_{nuc}$ in eq. (2.2) one of the
most efficient approaches, the multi configuration time-dependent
Hartree (MCTDH) method ~\cite{mctdh_1,mctdh_2,mctdh_3,mctdh_4,mctdh_5}
has been utilized. This method provides a very efficient algorithm
which can determine the quantal motion of the nuclei of a molecular
system evolving on one or several coupled electronic potential energy
surfaces. It has been applied many times with great success in the
last two decades.

For characterizing the vibrational degree of freedom R, we used FFT-DVR
(Fast Fourier Transformation-Discrete Variable Representation) with
$N{}_{R}$ basis elements distributed in the interval of $R=0.1-80\,a.u.$
The rotational degree of freedom $\theta$ was described by Legendre
polynomials \{P$_{J}$ (cos $\theta$)\}$_{j=0,1,2,...,N_{\theta}}$.
These so called primitive basis functions ($\chi$) were utilized
to represent the single particle functions ($\Phi$), which in turn
were used to expand the total nuclear wave function ($\psi$):

\[
\Phi_{j_{s}}^{(s)}(s,t)=\sum\displaylimits_{i=1}^{N_{s}}\ d_{j_{s}i}^{(s)}(t)\ \chi_{i}^{(s)}(s)\qquad\qquad s=R,\theta\tag{3.2.1}
\]

\[
\psi(R,\theta,t)=\sum\displaylimits_{j_{R}=1}^{n_{R}}\sum\displaylimits_{j_{\theta}=1}^{n_{\theta}}\ A_{j_{R}\ j_{\theta}}(t)\ \Phi_{j_{R}}^{(R)}(R,t)\ \Phi_{j_{\theta}}^{(\theta)}(\theta,t)\tag{3.2.2}
\]
\vspace{0.5cm}

The actual number of primitive basis functions in the calculations
were chosen to be $N_{R}=2048$ and $N_{\theta}=61$ for the vibrational
and rotational degrees of freedom, respectively. On both surfaces
and for both coordinates the same number of single particle functions
(SPF)s were applied. This number had to be raised as the intensity
and pulse length were increasing. Typical values were ranging from
$n_{R}=n_{\theta}=10...20$. Special attention has been payed to the
proper choice of basis so that convergence has been reached in each
propagation.\newpage{}

\section{Results and discussion \vspace{0.5cm}
}

Dynamical results eventuating from the numerical wave packet propagations
are presented in this section. In our numerical experiment the first
pump pulse ionizes the neutral D$_{2}$ molecule and generates a nuclear
wave packet on the ground electronic state of the molecular ion. A
second delayed probe pulse comes then to initiate the dissociation
process and leads to fragmentation of the D$_{2}^{+}$ molecular ion.
In the probe stage several different laser pulses - covering a wide
range of parameters - were used. We especially focus on the effect
of the time delay ($t_{delay}$) and the pulse duration $T$$_{p}$.
Above all, the impact of linear frequency chirp constitutes the main
subject of this study. All of our results are presented in terms of
total dissociation probability.

So as to better understand our dissociation yield results eventuating
from linearly chirped laser pulses (see Fig. 2 in \cite{csehi_2}),
we investigate now short pulse situations ($T{}_{p}$=$1\,fs$ and
$2\,fs$). Using shorter pulses the maximum value of the relative
chirp parameter (for $\beta=1$) corresponds to larger frequency changing
rates and, thus the position of the LICI moves faster. If the pulse
duration is just a few $fs$ then in our case the position of the
LICI can move with approximately the same speed as the wave packet
itself. This can help us to study in more detailed the impact of the
pulse chirping on observable quantities like the dissociation rate.
Moreover, if the pulse duration is much shorter than the period of
the field-free vibration of the wave packet, one can probe the time
delay dependence of the dissociation probability by getting rid of
the effects caused by the averaging over a long period of time and
on a wide range of internuclear distances and momenta. 

Our results for $T{}_{p}$=$1\,fs$ and $2\,fs$ are shown in Fig.
2 panels (A) and (C), respectively. Similarly to the situation of
longer pulses, the accounted dissociation rates depend strongly upon
the delay time $t_{delay}$. But for short pulses this delay time
dependence is not so smooth as has been found for longer pulses (see
Fig. 2 in \cite{csehi_2}). The curves display definite peaks at certain
time delays ca. $8\,fs$, $33\,fs$ and $57\,fs$. Fig. 2 panel (B)
shows that at these special delay times during its field-free motion
a major portion of the Franck Condon initialized wave packet is crossing
the LICI position ($R_{LICI}\sim2.9\,au.$ at $\lambda_{0}=200\,\mathrm{nm}$)
with a positive momentum\footnote{In the presence of the applied laser field the vibrational-rotational
wave packet can be considered very similar to the field-free one as
the small amount of dissociation yield that is provided by using short
laser pulse does not modify it significantly. Moreover up to a certain
delay time $t_{delay*}$the two wave packets are completely identical.
Minor deviations occur only for $t\gtrsim t_{delay*}$.}. By inspecting the differences between the negative and positive
chirp results (black dotted lines) it is obvious that the two dissociation
curves (dotted red and dashed blue) differ not only at the position
of the peaks, but almost in the whole delay time interval. The deviations
are significant in the vicinity of the peaks. At the peaks we can
notice a significant increased or decreased dissociation rate for
negatively or positively chirped pulses, respectively. This finding
could be attributed to the fact that for negative chirps the LICI
is moving together with the wave packet giving better chance for the
dissociation, while in case of positive chirps the opposite effect
is observed. However, a little bit away from the dissociation peaks
a reverse ordering of the negative and positive chirp dissociation
yields is found, namely positively chirped pulses provide larger dissociation
rates in those regions (see the negative peaks in the black dotted
lines around the positive peaks). The explanation of this finding
can be summarized as follows: For delay times that correspond to slightly
smaller or larger values than the temporal position of the dissociation
peaks, the LICI created by negatively chirped pulses is moving parallel
to the wave packet and they avoid each other (no overlap - smaller
yield). However, for positive chirps, the LICI is moving in the opposite
direction and at the beginning ($t_{delay}>t_{peak})$ or at the end
($t_{delay}<t_{peak})$ of the pulse the position of the LICI and
the outgoing wave packet still overlap to some extent giving larger
dissociation yield. The impact of the chirp is quite similar for $T$$_{p}$=$2\,fs$
(see panel C). 

By further increasing the $T$$_{p}$, one encounters a smooth change
in the behavior of the dissociation curves. Namely, the characteristic
peaks are smoothed out and the delay time dependence is suppressed
in the chirped case: the curves become more and more similar to the
$T_{p}=10\,fs$ situation (see Fig. 3 here and Fig. 2 in \cite{csehi_2}). 

\begin{figure}[p]
\centering{}\includegraphics[clip,width=7cm]{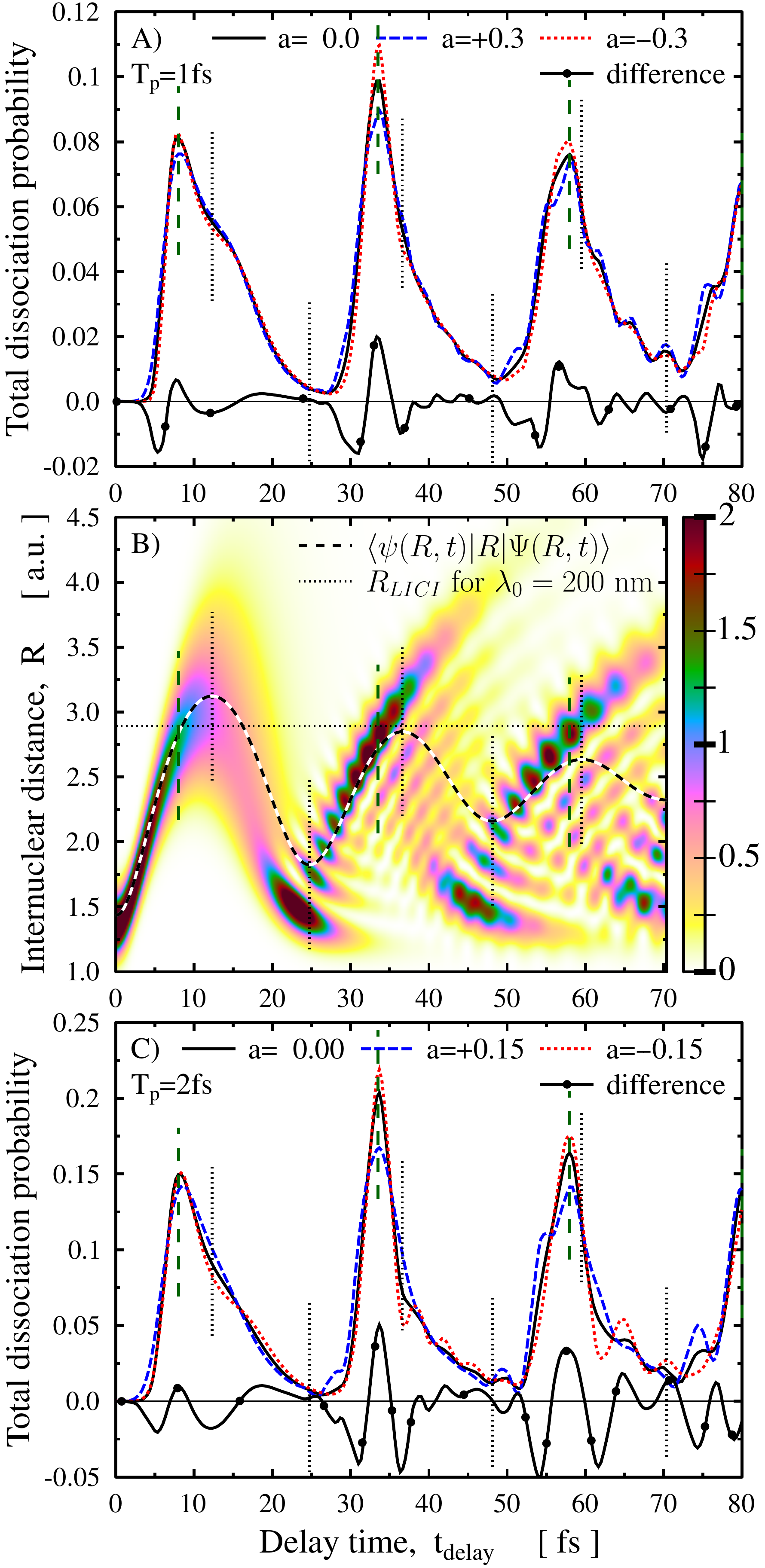} \caption{Dissociation probabilities of D$_{2}^{+}$ as a function of the delay
time $t_{delay}$ of the probe pulse. The peak intensity is $10^{13}W/cm^{2}$
and the central wavelength is $200\,nm$. All the panels include positive,
negative and TL chirp cases with the pulse duration being $1\,fs$
and $2\:fs$ (panels (A) and (C), respectively). In these panels the
black dotted curves present the difference between the dissociation
probabilities obtained by the positively and negatively chirped pulses.
Panel (B) shows the field-free vibration of the system. Here the square
of the vibrational wave function is represented by color code: the
darker the color the larger the amplitude of the wave packet. The
vertical dashed green and dotted black lines represent the position
of the local maxima of the dissociation probabilities obtained by
$1\,fs$ pulse and the extremal positions of the average internuclear
distances, respectively.}
\end{figure}

\begin{figure}[p]
\begin{centering}
\includegraphics[clip,width=6cm]{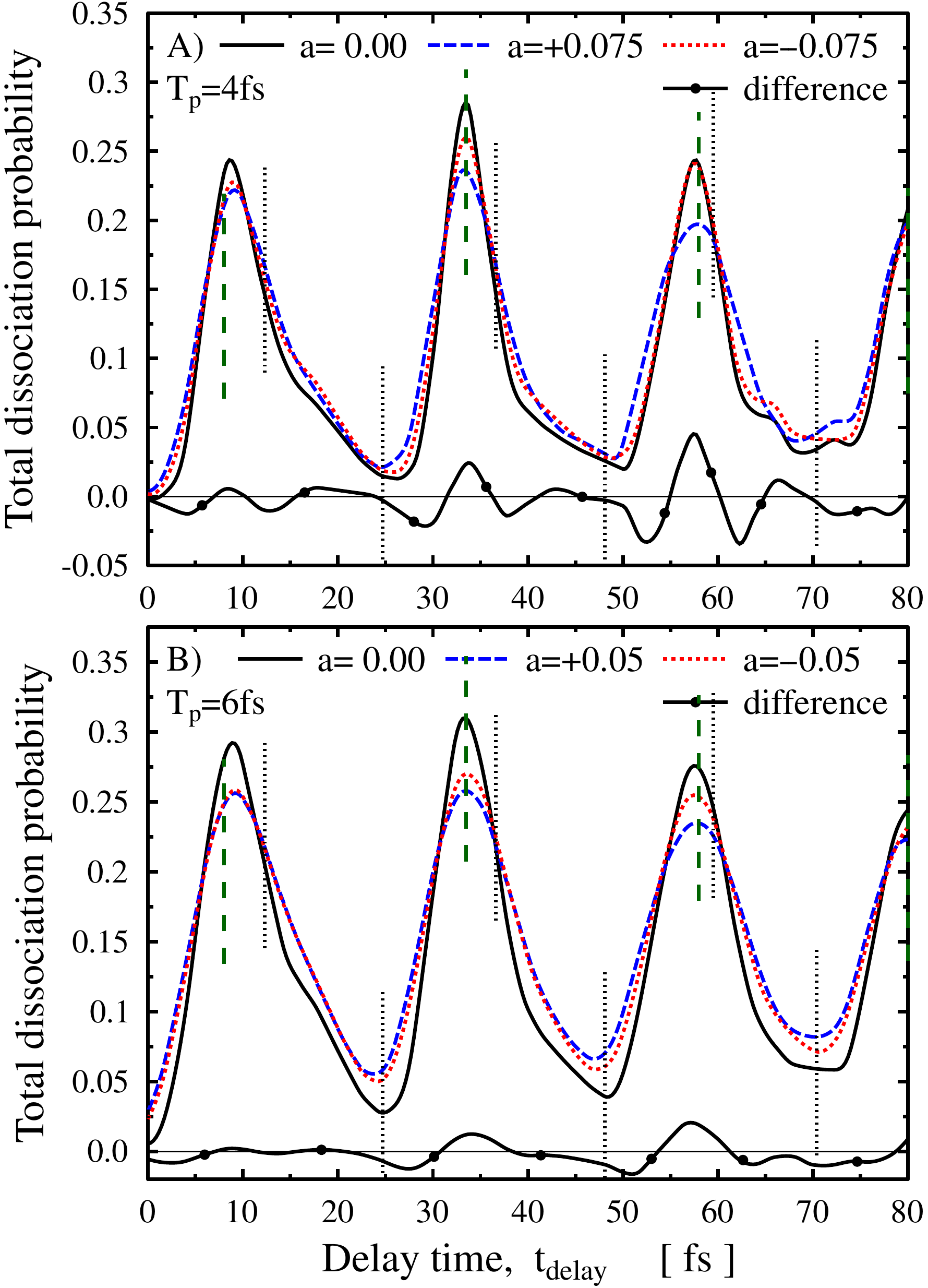}
\par\end{centering}

\caption{Dissociation probabilities of D$_{2}^{+}$ as a function of the delay
time $t_{delay}$ of the probe pulse. The peak intensity is $10^{13}W/cm^{2}$
and the central wavelength is $200\,nm$. Both panels include positive,
negative and TL chirp cases with the pulse duration being $4\,fs$
and $6\,fs$ (panels (A) and (B), respectively). The chirp parameters
were $a=\pm0.075$ and $a=\pm0.05$ for the $4\,fs$ and $6\,fs$
pulses, respectively. The black dotted curves present the difference
between the dissociation probabilities obtained by the positively
and negatively chirped pulses. The vertical dashed green and dotted
black lines represent the position of the local maxima of the dissociation
probabilities obtained by $1\,fs$ pulse and the extremal positions
of the average internuclear distances, respectively. }
\end{figure}

The impact of the chirp is further analyzed in Fig. 4 for $T$$_{p}$=$1$$fs$
pulse length. In this figure the dissociation probability is plotted
against the chirp parameter $a$, ranging from $a$$=-0.3$ to $a$$=0.3$
($\beta\approx\mp1$) at different delay times around one of the peaks
($t_{delay}=32\ldots35$ $fs$). Among the studied delay times the
dissociation have the largest rate at $t_{delay}=33.5\,fs$ in the
whole inspected parameter range of the chirp. At this special time
delay the total dissociation rate depends linearly upon the chirp
parameter $a$. Negative values increase the dissociation rate while
the positive chirp suppresses it. Changing the delay time, not only
the dissociation rate decreases but more and more pronounced quadratic
dependence on the chirp parameter can be observed -- especially for
shorter delay times. As a result of this quadratic dependence, even
the negatively chirped pulses provide less dissociation rate than
the transform limited one at $t_{delay}=32\,fs$. 

The outgoing wave packet has a large amplitude between the $30\,fs$
and $36\,fs$ time delays in which time interval the wave packet of
the field-free motion strongly overlaps with the $R{}_{LICI}$ region.
In the above time interval the wave packet is moving outwards, almost
reaches its maximal $R$ value and its momentum is decreasing towards
0 but it still remains positive. Therefore, applying a sharp pulse
that is centered in the $30\,fs$ -$36\,fs$ interval one can achieve
relatively significant population transfer since the density is big
and at the same time its major part is in the $R{}_{LICI}$ region.

\begin{figure}[p]
\includegraphics[clip,width=6cm]{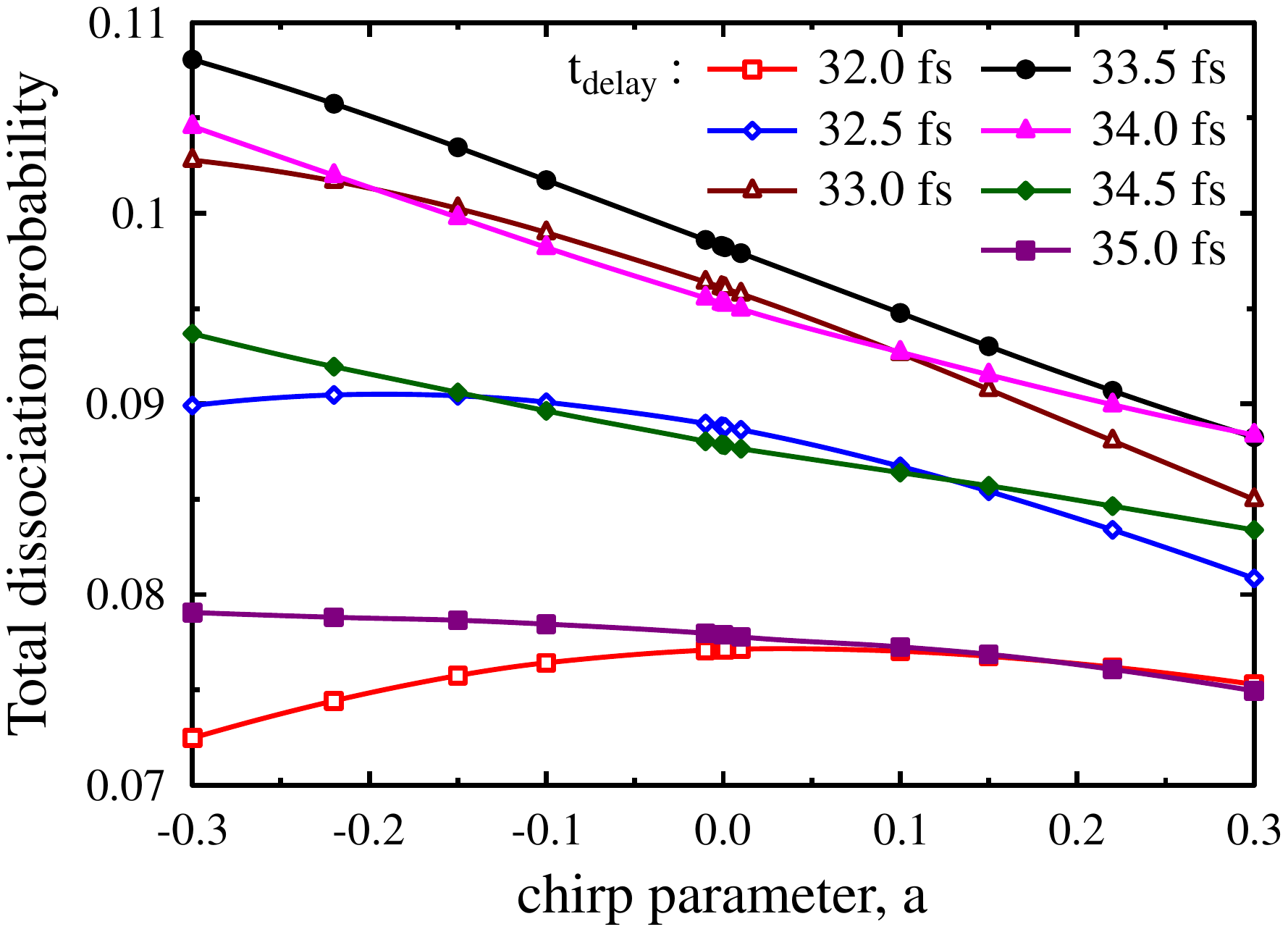} \caption{Dissociation probabilities of D$_{2}^{+}$ as a function of the chirp
parameter around $t$$_{delay}$=$33.5\,fs$ for a peak intensity
of $10^{13}W/cm^{2}$ . The central wavelength is $200\,nm$ and $T_{p}=1\,fs$.}
\end{figure}


By chirping the frequency different properties of the pulse can change.
Although we preserved the spectra of the pulse and it remained unchanged
during the simulations, the effective pulse duration is increased
and the effective intensity is decreased by a same factor of $\sqrt{1+\beta^{2}}$.
As we were interested in using as large frequency changing rate as
possible, we have applied pulses with $\beta$ very close to one,
where the widening of the pulses are around $\sqrt{2}$ , which -
beside the frequency chirping - can also provide a significant effect
on the dissociation process. To understand clearly the effect of the
pulse widening on the behavior of the dissociation curves obtained
for longer pulse durations (see on Fig. 2 in \cite{csehi_2}) we performed
calculations by using broadened TL pulses and compared the results
obtained to the $T$$_{p}$=$10$$fs$ length chirped ones. Obtained
results by this artificially broadened TL pulse ($T{}_{p}$=$14.1\,fs$)
-- despite of the narrower spectra in this case -- for the dissociation
probabilities were between the original results for $T$$_{p}$=$10\,fs$
with positively and negatively chirped pulses. This implies that the
suppression of the dissociation yields emerges from the broadening
of the chirped pulses and the linearly changing frequency is responsible
only for the difference between the dissociation rates belonging to
the two kind of chirped pulses. 

\begin{figure}[p]
\includegraphics[clip,width=6cm]{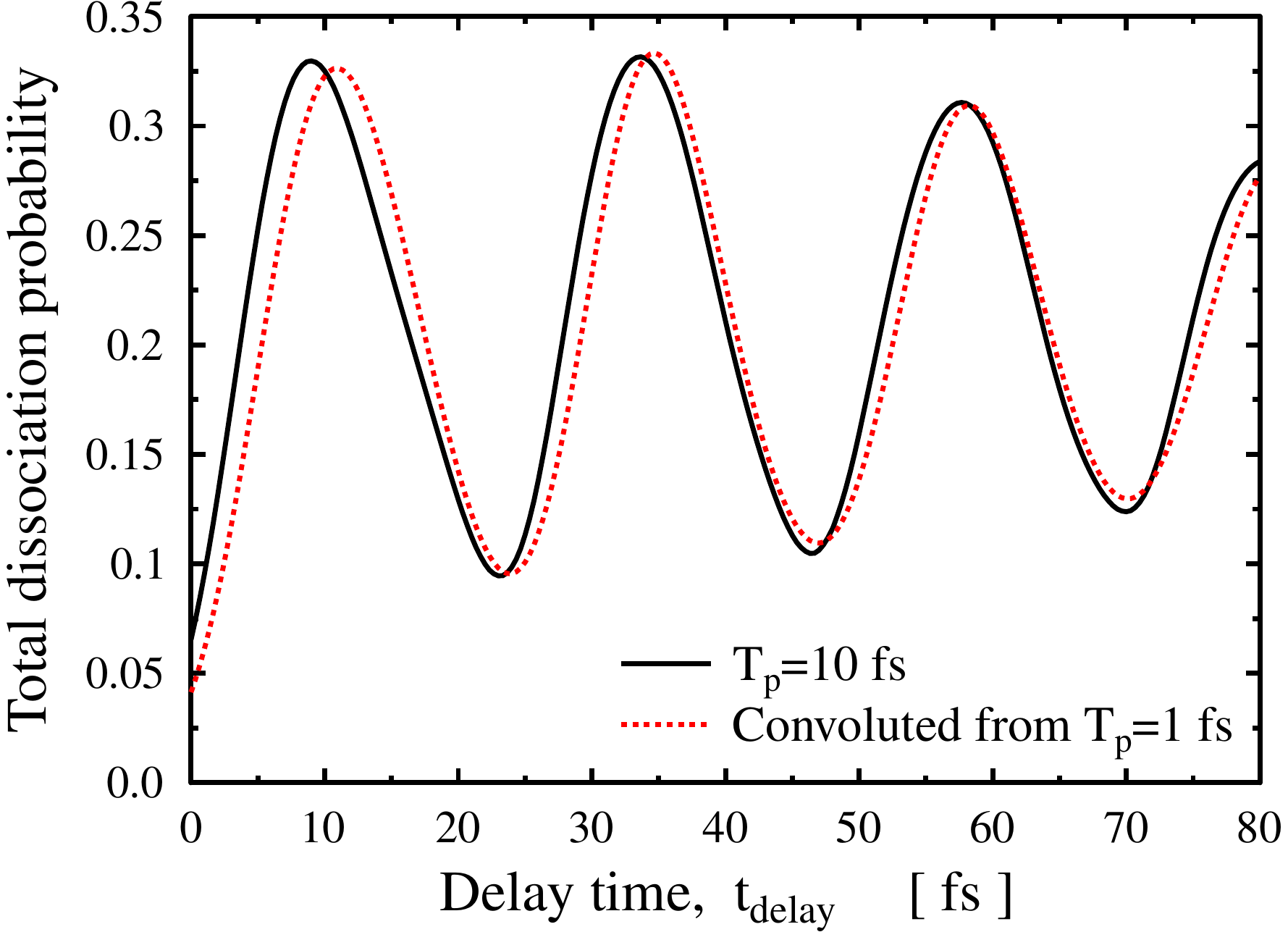} \caption{Comparison of dissociation probabilities for D$_{2}^{+}$. Black line:
results of wave packet propagation ($T$$_{p}$ = $10\,fs$) ; red
line: values calculated according to eq. (4.1) obtained by weighting
the $T{}_{p}$ = $1\,fs$ probabilities. }
\end{figure}

To check the validity of this explanation for the TL pulses we constructed
a model for calculating the dissociation yields for wide pulses with
the help of the dissociation yields obtained by using short pulses.
In this picture it is assumed that the temporal dissociation rate
during the presence of the long pulse is proportional to the temporal
intensity and to the calculated total dissociation rate obtained by
short pulse. Therefore the dissociation probability with wide pulse
(eg. $10\,fs$) at $t_{delay}$ delay time is obtained by the integrated
values of the dissociation probabilities with short pulse (eg. $1\,fs$)
weighted by the envelope of the given pulse. The respective expression
for the approximate dissociation rate can be written as a convolution
and reads: 

\begin{align*}
P_{diss}^{model}(t_{delay},T_{p}^{wide})=scale*\int_{0}^{\infty}P_{diss}(t,T_{p}^{short})G(t-t_{delay})dt\quad\tag{4.1}
\end{align*}

\vspace{0.5cm}
 where $P_{diss}(t,T_{p}^{short})$ is the dissociation rate for short
pulses with $t_{delay}=t$. The convolution function $G(t-t_{delay})$
is the temporal intensity of the Gaussian pulse ,i.e., $G(t-t_{delay})=e^{-4log2\left(\frac{\Delta t}{T_{p}^{wide}}\right)^{2}}$,
and the $scale$ was chosen to fit the model to real dissociation
rate for the situation of wide pulse at a given delay time. 

Results of the numerical simulations obtained from the above model
are demonstrated in Fig. 5 . After scaling the numerical values obtained
from eq. (4.1) so that at a certain point they coincide with the original
curve, obtained by the $10\,fs$ pulse calculations, a very good agreement
was observed not only for $T_{p}^{short}$=$1\,fs$, but also for
$T_{p}^{short}$=$2\,fs$ pulses. This agreement perfectly supports
our finding, namely that the suppression of the dissociation rate
provided by the chirped laser pulses \cite{csehi_2}, is the effect
of the effective pulse widening. When the $t_{delay}$ is in the vicinity
of the peaks seen in Fig. 2 then the reduced intensity of the pulse,
which is a side effect of the pulse widening, provides a smaller total
dissociation yield for the chirped pulse. On the other hand, if the
pulse is centered in between two peaks, a wider pulse -- despite of
the reduced effective intensity -- will have a larger temporal intensity
at that times when the dissociation can take place with the largest
rates.

\newpage{}

\section{Conclusions \vspace{0.5cm}
}

Linearly varying frequency chirped laser fields of different pulse
durations have been applied in the present paper for studying the
photodissociation dynamics of D$_{2}^{+}$. In our treatment the effects
of positive, negative and zero chirps have been explored in terms
of total dissociation probabilities as a function of delay time between
the ionization of D$_{2}$ and the maximum of the probe pulse ($t_{delay}$).

Our study was carried out in the recently introduced LICI framework
in which the created light-induced conical intersection is present
as long as the laser light is switched on so that the molecule can
rotate with respect to the field. 

The present paper is a comprehensive study in the sense that besides
the effect of chirping, two other laser field parameter dependence,
such as $T_{p}$ and $t_{delay}$ have been investigated. 

One of our most important finding is that when the oscillating nuclear
wave packet reaches its maximal position along the internuclear coordinate
$R$ and its momentum is zero, we were able to fine tune the photodissociation
probability. It was demonstrated that when the time delay of the pulse
fits to the wave packet being in the above region the use of negatively
chirped pulses increases the dissociation probability, while the opposite
is true for the positively chirped pulses. Although this observation
is not momentous, it is fully consistent with our initial anticipation
that the frequency chirping of a pulse results in the movement of
the LICI and that this LICI then has a considerable impact on the
dynamics. 

To continue, we plan to go much beyond and perform calculations using
chirped laser pulses in which the time dependence of the laser frequency
is designed so as to achieve a desired outcome of a given molecular
reaction. Our future prospect is to understand better the relation
between the evolving wave packet and the frequency chirping. Being
able to directly follow the time evolution of the nuclear wave packet
density is expected to open up the possibility to control the dissociation
probability or any other dynamical properties of the system under
investigation.

\begin{acknowledgments}
The authors acknowledge the financial support by the Deutsche Forschungsgemeinschaft
(Project ID CE10/50-2). Á.V. also acknowledges the OTKA (NN103251)
project. For this work, the supercomputing service of NIIF has been
used.
\end{acknowledgments}






\begin{lyxgreyedout}
\end{lyxgreyedout}


\begin{thebibliography}{10}
\bibitem{born1} M. Born, R. Oppenheimer, Ann. Phys. \textbf{84},
457 (1927). 

\bibitem{horst} H. Köppel, W. Domcke, L. S. Cederbaum, Adv. Chem.
Phys. \textbf{57}, 59 (1984). 

\bibitem{baer} M. Baer, Phys. Rep. \textbf{358}, 75 (2002). 

\bibitem{graham} G. A. Worth, L. S. Cederbaum, Annu. Rev. Phys. Chem.
\textbf{55}, 127 (2004). 

\bibitem{domcke} Conical Intersections: Electronic Structure, Dynamics
and Spectroscopy, edited by W. Domcke, D. R. Yarkony, H. Köppel (World
Scientific: Singapore, 2004). 

\bibitem{baer2} M. Baer, Beyond Born Oppenheimer: Electronic Non-Adiabatic
Coupling Terms and Conical Intersections (Wiley: Hoboken, NJ, 2006). 

\bibitem{matsika} S. Matsika, Rev. Comput. Chem., \textbf{23}, 83
(2007). 

\bibitem{althorpe} S. C. Althorpe, T. Stecher and F. Bouakline, J.
Chem. Phys. \textbf{129}, 214117 (2008). 

\bibitem{bouakline} F. Bouakline, Chem. Phys. \textbf{442}, 31 (2014).



\bibitem{domcke2} W. Domcke, D. R. Yarkony, Annu. Rev. Phys. Chem.
\textbf{63}, 325 (2012). (and references therein) 

\bibitem{olivucci} P. El-Khoury, I. Schapiro, M. Huntress, F. Melaccio,
S. Gozem, L. M. Frutos and M. Olivucci, in CRC Handbook of Organic
Photochemistry and Photobiology. 3rd Edition (eds A. Griesbeck and
F. Ghetti) 1029-1056 (CRC Press, Boca Raton, FL, 2012). (and references
therein) 

\bibitem{robb} H. H. Fielding, M. A. Robb, Phys. Chem. Chem. Phys.
\textbf{12}, 15569 (2010).



\bibitem{olivucci2} S. Gozem, A. Krylov and M. Olivucci, J. Chem.
Theo. Comp. \textbf{9}, 284 (2012). 

\bibitem{martinez2} T. J. Martinez, Nature \textbf{467}, 412 (2010). 

\bibitem{sobolewski} S. Yamazaki, W. Domcke and A. L. Sobolewski,
J. Phys. Chem. A \textbf{112}, 11965 (2008).



\bibitem{nimrod} N. Moiseyev, M. Sindelka and L. S. Cederbaum, J.
Phys. B \textbf{41}, 221001 (2008). 

\bibitem{milan} M. Sindelka, N. Moiseyev and L. S. Cederbaum, J.
Phys. B \textbf{44}, 045603 (2011). 

\bibitem{gabor_1} G. J. Halász, Á. Vibók, M. Sindelka, N. Moiseyev
and L. S. Cederbaum, J. Phys. B \textbf{44}, 175102 (2011). 

\bibitem{gabor_2} G. J. Halász, M. Sindelka, N. Moiseyev, L. S. Cederbaum
and Á. Vibók, J. Phys. Chem. A \textbf{116}, 2636 (2012). 

\bibitem{gabor_3} G. J. Halász, Á. Vibók, M. Sindelka, L. S. Cederbaum
and N. Moiseyev, Chem. Phys. \textbf{399}, 146 (2012). 

\bibitem{gabor_4} G. J. Halász, Á. Vibók, N. Moiseyev and L. S. Cederbaum,
J. Phys. B \textbf{45}, 135101 (2012). 

\bibitem{chiang} L. S. Cederbaum, Y. C. Chiang, P. V. Demekhin and
N. Moiseyev, Phys. Rev. Lett. \textbf{106}, 123001 (2011). 

\bibitem{pawlak} M. Pawlak and N. Moiseyev, Phys. Rev. A \textbf{92},
023403 (2015).

\bibitem{kim} J. Kim, H. Tao, J. L. White, V. S. Petrovi, T. J. Martinez
and P. H. Bucksbaum, J. Phys. Chem. A \textbf{116}, 2758 (2012). 

\bibitem{philip} P. V. Demekhin and L. S. Cederbaum, J. Chem. Phys.
\textbf{139}, 154314 (2013). 

\bibitem{ware} M. Ware, A. Natan, and P. Bucksbaum, Experimental
evidence of light induced conical intersection in photodissociation
of diatomic molecules, Bulletin of the American Physical Society \textbf{59},
(2014). 

\bibitem{corrales} M. E. Corrales, J. González-Vázquez, G. Balerdi,
I. R. Solá, R. de Nalda and L. Bañares, Nat. Chem. \textbf{6}, 785
(2014). 

\bibitem{ware2} A. Natan, M. R. Ware and P. H. Bucksbaum, Ultrafast
Phenomena XIX, Springer Proceedings in Physics \textbf{162}, 122 (2015). 

\bibitem{sola} I. R. Solá, J. González-Vázquez, R. de Nalda, and
L. Bañares, Phys. Chem. Chem. Phys. \textbf{17}, 13183 (2015). 

\bibitem{martinez} J. Kim, H. Tao, T. J. Martinez and P. H. Bucksbaum,
J. Phys. B \textbf{48}, 164003 (2015).

\bibitem{gabor_5} G. J. Halász, Á. Vibók, H. D. Meyer and L. S. Cederbaum,
J. Phys. Chem. A \textbf{117}, 8528 (2013). 

\bibitem{gabor_6} G. J. Halász, Á. Vibók, N. Moiseyev and L. S. Cederbaum,
Phys. Rev. A \textbf{88}, 043413 (2013). 

\bibitem{csehi} G. J. Halász, A. Csehi, Á. Vibók and L. S. Cederbaum,
J. Phys. Chem. A \textbf{118}, 11908 (2014). 

\bibitem{gabor_7} G. J. Halász, Á. Vibók and L. S. Cederbaum, J.
Phys. Chem. Lett. \textbf{6}, 348 (2015). 

\bibitem{csehi_2} A. Csehi, G. J. Halász, L. S. Cederbaum and Á.
Vibók, J. Chem. Phys. \textbf{143}, 014305 (2015). 

\bibitem{csehi_3} G. J. Halász, A. Csehi and Á. Vibók, Theor. Chem.
Acc. \textbf{134}, 128 (2015).



\bibitem{bandrauk_1} A. D. Bandrauk and M. Sink, Chem. Phys. Lett.
\textbf{57}, 569 (1978). 

\bibitem{bandrauk_2} A. D. Bandrauk and M. Sink, J. Chem. Phys. \textbf{74},
1110 (1981). 

\bibitem{zavriyev} A. Zavriyev, P. H. Bucksbaum, H. G. Muller and
D. V. Schumacher, Phys. Rev. A \textbf{42}, 5500 (1990). 

\bibitem{bandrauk_3} E. E. Aubanel, J. M. Gauthier and A. D. Bandrauk,
Phys. Rev. A \textbf{48}, 2145 (1993). 

\bibitem{charron_1} E. Charron, A. Giusti-Suzor and F. H. Mies, Phys.
Rev. A \textbf{49}, R641 (1994). 

\bibitem{atabek_1} S. Chelkowski, T. Zuo, O. Atabek and A. D. Bandrauk,
Phys. Rev. A \textbf{52}, 2977 (1995). 

\bibitem{charron_2} A. Giusti-Suzor, F. H. Mies, L. F. DiMauro, E.
Charron and B. Yang, J. Phys. B \textbf{28}, 309 (1995). 

\bibitem{atabek_2} R. Numico, A. Keller and O. Atabek, Phys. Rev.
A \textbf{52}, 1298 (1995). 

\bibitem{sanchez} I. Sánchez and F. Martín, Phys. Rev. A \textbf{57},
1006 (1998). 

\bibitem{sandig} K. Sandig, H. Figger and T. V. Hansch, Phys. Rev.
Lett. \textbf{85}, 4876 (2000). 

\bibitem{atabek_3} V. N. Serov, A. Keller, O. Atabek and N. Billy,
Phys. Rev. A \textbf{68}, 053401-1 (2003). 

\bibitem{posthumus} J. H. Posthumus, Rep. Prog. Phys. \textbf{67},
623 (2004). 

\bibitem{atabek_4} V. N. Serov, A. Keller, O. Atabek, H. Figger and
D. Pavidic, Phys. Rev. A \textbf{72}, 033413 (2005). 

\bibitem{uhlmann} M. Uhlmann, T. Kunert and R. Schmidt, Phys. Rev.
A \textbf{72}, 045402-1 (2005). 

\bibitem{wang} P. Q. Wang, A. M. Sayler, K. D. Carnes, J. F. Xia,
M. A. Smith, B. D. Esry and I. Ben-Itzhak, Phys. Rev. A \textbf{74},
043411-1 (2006). 

\bibitem{anis_1} F. Anis and B. D. Esry, Phys. Rev. A \textbf{77},
033416-1 (2008). 

\bibitem{anis_2} F. Anis, T. Cackowski, B. D. Esry, J. Phys. B \textbf{42},
091001-1 (2009). 

\bibitem{hua} J. J. Hua and B. D. Esry, Phys. Rev. A \textbf{80},
013413 (2009). 

\bibitem{satrajit_1} S. Adhikari, A. K. Paul, D. Mukhopadhyay, G.
J. Halász, Á. Vibók, R. Baer and M. Baer, J. Phys. Chem. A \textbf{113},
7331 (2009). 

\bibitem{satrajit_2} A. K. Paul, S. Adhikari, M. Baer and R. Baer,
Phys. Rev. A \textbf{81}, 013412-1 (2010). 

\bibitem{calvert} C. R. Calvert, W. A. Bryan, W. R. Newell and I.
D. Williams, Phys. Rep. \textbf{491}, 1 (2010). 

\bibitem{thumm} U. Thumm, T. Niederhausen and B. Feuerstein, B. Phys.
Rev. A \textbf{77}, 063401-1 (2008). 

\bibitem{fischer_1} M. Fischer, F. Grossmann, R. Schmidt, J. Handt,
S. M. Krause and J. M. Rost, New. J. Phys. \textbf{13}, 053019-1 (2011). 

\bibitem{fischer_2} M. Fischer, U. Lorenz, B. Schmidt and R. Schmidt,
Phys. Rev. A \textbf{84}, 033422-1 (2011). 

\bibitem{mckenna} J. McKenna, F. Anis, A. M. Sayler, B. Gaire, N.
G. Johnson, E. Parke, K. D. Carnes, B. D. Esry and I. Ben-Itzhak,
Phys. Rev. A \textbf{85}, 023405-1 (2012). 

\bibitem{he} H. X. He, R. F. Lu, P. Y. Zhang, K. L. Han and G. Z.
He, J. Chem. Phys. \textbf{136}, 024311-1 (2012). 

\bibitem{furukawa} Y. Furukawa, Y. Nabekawa et al., Opt. Lett. \textbf{37},
2922 (2012). 

\bibitem{fischer_3} J. Handt, S. M. Krause, J. M. Rost, M. Fischer,
F. Grossmann and R. Schmidt, arXiv 2011, arXiv:1103.1565v2.



\bibitem{igarashi} A. Igarashi, Eur. Phys. J. D, \textbf{68}, 266
(2014). 

\bibitem{haxton2} D. J. Haxton, Phys. Rev. A \textbf{88}, 013415
(2013). 

\bibitem{haxton} D. J. Haxton, K. V. Lawler and C. W. McCurdy, Phys.
Rev. A \textbf{91}, 062502 (2015). 











\bibitem{krause} V.S. Malinovsky and J.L. Krause, Eur. Phys. J. D
\textbf{14}, 147 (2001). 

\bibitem{datta} A. Datta, S. S. Bhattacharyya and B. Kim, Phys. Rev.
A \textbf{65}, 043404 (2002). 

\bibitem{kosloff} G. Katz, M. A. Ratner and R. Kosloff, New J. Phys.
\textbf{12}, 015003 (2010). 

\bibitem{prabhudesai} V. S. Prabhudesai, U. Lev, A. Natan, B. D.
Bruner, A. Diner, O. Heber, D. Strasser, D. Schwalm, I. Ben-Itzhak,
J. J. Hua, B. D. Esry, Y. Silberberg, D. Zajfman, Phys. Rev. A \textbf{81},
023401 (2010). 

\bibitem{natan} A. Natan, U. Lev, V. S. Prabhudesai, B. D. Bruner,
D. Strasser, D. Schwalm, I. Ben-Itzhak, O. Heber, D. Zajfman and Y.
Silberberg, Phys. Rev. A \textbf{86}, 043418 (2012). 

\bibitem{prabhudesai2} V. S. Prabhudesai, A. Natan, B. D. Bruner,
Y. Silberberg, U. Lev, O. Heber, D. Strasser, D. Schwalm, D. Zajfman,
I. Ben-Itzhak, J. Kor. Phys. Soc. \textbf{59}, 2890 (2011). 

\bibitem{chang} B. Y. Chang, S. Shin, J. Santamaria and I. R. Sola,
J. Phys. Chem. A \textbf{116}, 2691 (2012). 

\bibitem{forre} S. Askeland and M. Førre, Phys. Rev. A \textbf{88},
043411 (2013). 

\bibitem{zhang} C. P. Zhang, X. Y. Miao, Spect. Lett. \textbf{47},
267 (2014). 

\bibitem{Natan} U. Lev, L. Graham, C. B. Madsen, I. Ben-Itzhak, B.
D. Bruner. B. D. Esry, H. Frosting, O. Heber, A. Natan and V. S. Prabhudesai,
J. Phys. B \textbf{48}, 201001 (2015).

\bibitem{Novitsky} D. V. Novitsky, Optic Comm. \textbf{358}, 202
(2016).

\bibitem{v1v2} F. V. Bunkin and I. I. Tugov, Phys. Rev. A \textbf{8},
601 (1973). 

\bibitem{tdm} S. I. Chu, C. Laughlin, and K. Datta, Chem. Phys. Lett.
\textbf{98}, 476 (1983).

\bibitem{Chu} S. I. Chu, J. Chem. Phys. \textbf{75}, 2215 (1981).

\bibitem{pulse} D. B. Milosevic, G. G. Paulus, D. Bauer, and W. Becker,
J. Phys. B \textbf{39}, R203 (2006).



\bibitem{mctdh_1} H. D. Meyer, U. Manthe and L. S. Cederbaum, Chem.
Phys. Lett. \textbf{165}, 73 (1990). 

\bibitem{mctdh_2} U. Manthe, H. D. Meyer and L. S. Cederbaum, J.
Chem. Phys. \textbf{97}, 3199 (1992). 

\bibitem{mctdh_3} M. H. Beck, A. Jäckle, G. A. Worth and H. D. Meyer,
Phys. Rep. \textbf{324}, 1 (2000). 

\bibitem{mctdh_4} G. A. Worth et al. The MCTDH package, version 8.2;
University of Heidelberg: Heidelberg, Germany, 2000. H. D. Meyer et
al. The MCTDH package, versions 8.3 and 8.4; University of Heidelberg,
Germany, 2002 and 2007. http://mctdh.uni-hd.de/. 

\bibitem{mctdh_5} Multidimensional Quantum Dynamics: MCTDH Theory
and Applications, edited by H. D. Meyer, F. Gatti, G. A. Worth, (Wiley-VCH:
Weinheim, 2009).

\end{thebibliography}
\end{document}